\documentclass[pra,twocolumn]{revtex4}
\usepackage{graphicx}
\begin{document}

\bibliographystyle{unsrt}

\date{May 28, 2016}

\title{Nanofiber-segment ring resonator}
\author{D.E. Jones}
\author{G.T. Hickman}
\author{J.D. Franson}
\author{T.B. Pittman}
\email{todd.pittman@umbc.edu}
\affiliation{Physics Department, University of Maryland Baltimore
County, Baltimore, MD 21250}

\begin{abstract}

We describe a fiber ring resonator comprised of a relatively long loop of standard single-mode fiber with a short nanofiber segment. The evanescent mode of the nanofiber segment allows the cavity-enhanced  field to interact with atoms in close proximity to the nanofiber surface.  We report on an experiment using a warm atomic vapor and low-finesse cavity, and briefly discuss the potential for reaching the strong coupling regime of cavity QED by using trapped atoms and a high-finesse cavity of this kind.

\end{abstract}


\maketitle

A compelling vision of the quantum internet involves atom-cavity nodes that are linked by flying photons propagating through fiber channels \cite{kimble2008}.  Within this context, there is a need for highly efficent coupling between the fiber mode and the cavity mode \cite{spillane2003,aoki2006,tiecke2014}, or the development of ``all-fiber'' cavities in which this coupling is inherent \cite{muller2010,hunger2010}. The recent development of an all-fiber cavity formed by two fiber Bragg gratings (FBGs) enclosing the nanofiber waist of a tapered optical fiber (TOF) is a particularly exciting prospect \cite{wuttke2012}. In that system, the combination of the tight transverse confinement of the nanofiber's evanescent optical mode with the longitudinal confinement of the Fabry-Perot geometry can lead to very strong  interactions with atoms in the cavity mode. In fact, this type of nanofiber-based Fabry-Perot  cavity has recently been used to reach the strong coupling regime of cavity QED \cite{kato2015}.

In this paper, we describe a closely related nanofiber-based cavity that uses the geometry of a conventional fiber-ring resonator \cite{heebner2004} rather than a Fabry-Perot cavity. When atoms are present in the nanofiber evanescent mode, this can essentially be viewed as a nonlinear fiber ring resonator \cite{heebner1999} where the bulk material nonlinearity of the silica fiber has been replaced with a resonant atomic interaction. We experimentally demonstrate this nonlinearity by saturating the atomic response with cavity fields at ultralow input power levels. We then briefly discuss the possibility of using this nanofiber-based ring resonator geometry for cavity QED experiments in analogy with \cite{kato2015}.

\begin{figure}[t]
\includegraphics[width=3.25in]{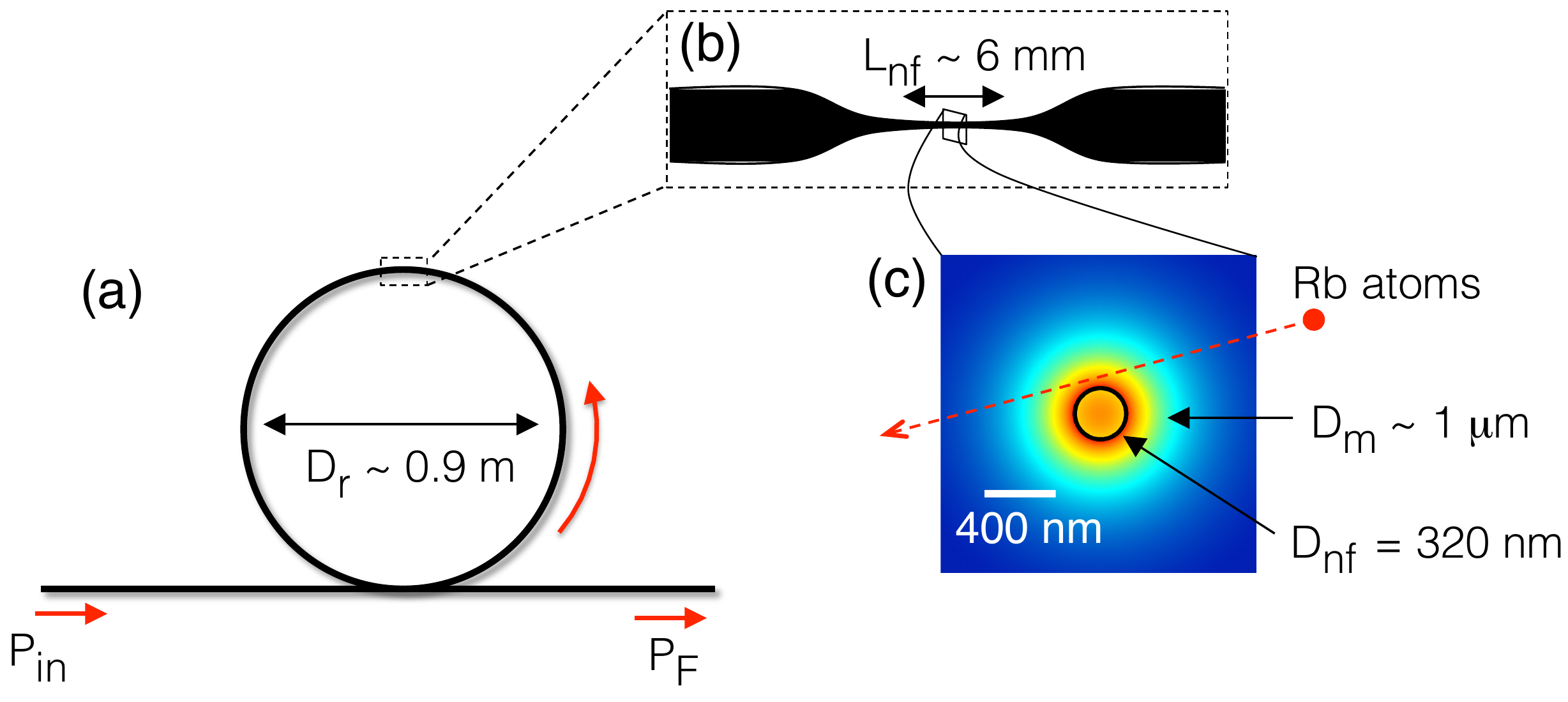}
\caption{(Color online) (a) overview of the nanofiber-segment (NFS) ring resonator geometry.  (b) the dashed box shows a zoom-in on the tapered nanofiber segment. (c) color contour plot of the calculated nanofiber mode in the transverse plane \protect\cite{kien2004}. The nanofiber diameter (black circle) is $D_{nf} \sim 320$ nm. The mode field diameter is $D_{m} \sim 1 \:  \mu$m.  Thermal Rb atoms fly through this mode on a timescale of a few ns. }
\label{fig:fig1}
\end{figure}

An overview of the system is shown in Figure \ref{fig:fig1}. A small region at the top of a standard single-mode fiber ring resonator is tapered to provide a short sub-wavelength diameter nanofiber segment. The evanscent mode guided by this nanofiber segment interacts with surrounding atoms \cite{spillane2008,nieddu2016}, and the cavity resonance is tuned to match the atomic resonance by adjusting the overall diameter of the ring.  Note that this geometry is fundamentally different than microfiber loop  (MFL) resonators in which the entire ring (including the coupling region) is comprised of micro- or nanofiber \cite{sumetsky2005}. MFL-type resonators have the advantage of very small mode volumes (loop diameters), but typically suffer from low finesse due to difficulty in controlling the coupling region \cite{xiao2011}. In contrast, the nanofiber-segment (NFS) ring resonator in Figure \ref{fig:fig1} has a comparatively large mode volume (ring diameter), but the possibility of  very high finesse by utilizing mature large-scale  fiber coupler technologies \cite{wang2009}.

In this initial experimental work we use a warm vapor of rubidium atoms (T $\sim 85^{o}$C), and choose a relatively large overall ring diameter of $D_{r} \sim 0.9$ m. This gives a short free spectral range (FSR $\sim 76$ MHz) that provides several cavity resonances within the Doppler broadened Rb linewidth ( $\sim 750$ MHz). The cavity response (both with and without the atoms) is then measured by scanning a tunable narrowband diode laser (Newport Velocity TLB-6700) across several FSRs.  The nanofiber segment has a diameter of roughly $D_{nf} \sim 320$ nm and a length of $L_{nf}\sim 6$ mm \cite{jones2014}. For Rb resonant light at 780 nm, this $D_{nf}$ guides an evanscent mode with a diameter of $D_{m} \sim 1 \:  \mu$m.

\begin{figure}[t]
\includegraphics[width=3.25in]{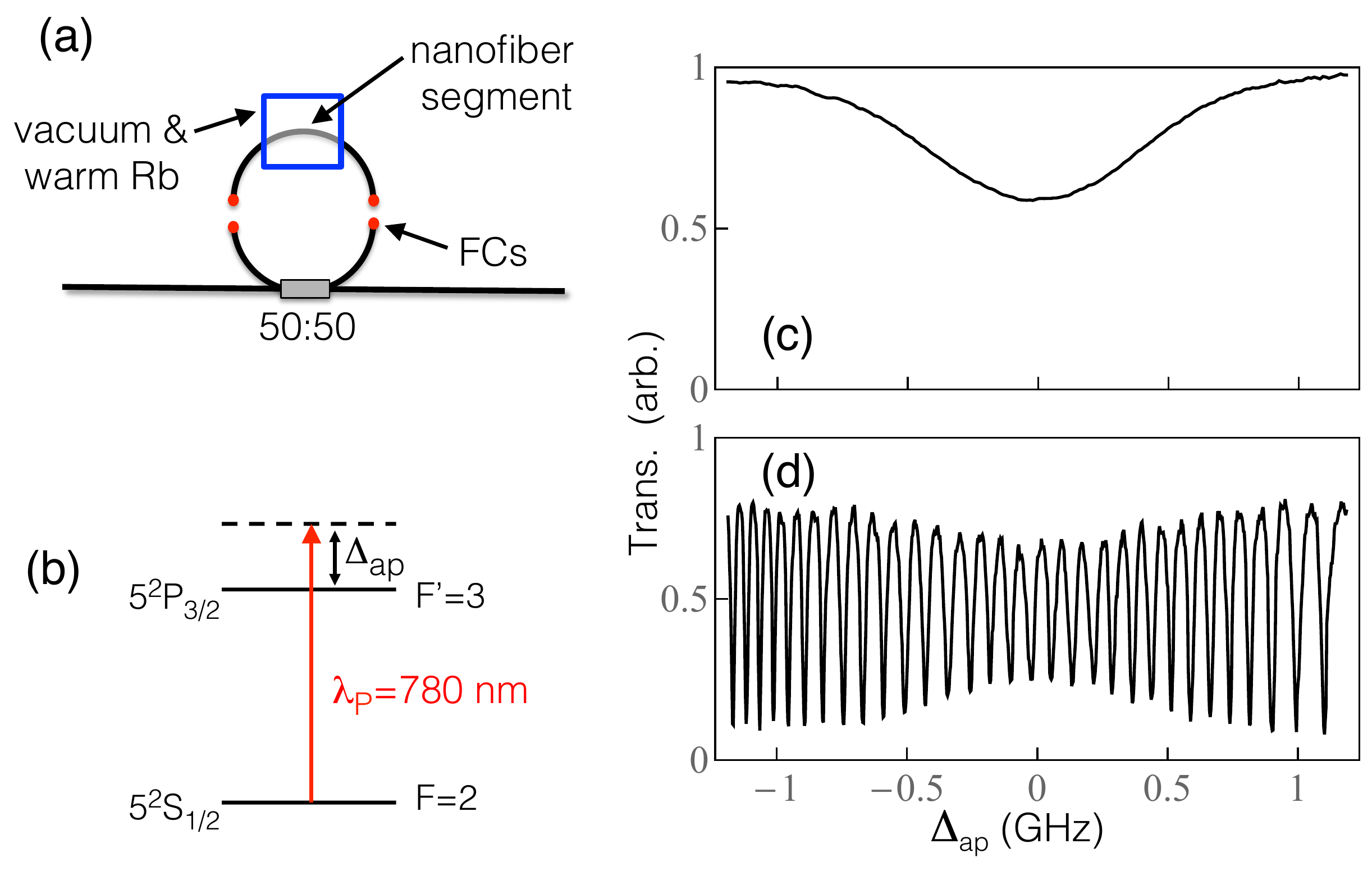}
\vspace*{-.1in}
\caption{(Color online) (a) overview of the experiment. Fiber connectors (FC's) allows a quick change between single-pass measurements through the nanofiber and measurements using the full NFS ring cavity geometry. (b) relevant $^{85}$Rb energy level diagram. (c) transmission spectrum for a single-pass through the nanofiber. (d) analagous transmission spectrum through the full NFS ring cavity geometry. The cavity has a FSR $\sim 76$ MHz and  $\mathcal{F} \sim 3.6$. Rb absorption provides an extra internal cavity loss mechanism that moves the cavity from critical coupling into the strongly undercoupled regime near $\Delta_{ap} =0$.}
\label{fig:fig2}
\end{figure}

In a ring resonator geometry, critical coupling is achieved when the coupling loss into the ring is equal to the total internal loss  \cite{yariv2000}. For the NFS ring resonator studied here, the intrinsic internal loss is roughly $50 \%$. This is primarily due to an overall initial TOF transmission of $70 \%$ (a non-adiabatic taper) and a subsequent degradation due to Rb accumulation on the nanofiber surface \cite{lai2013}. As shown in Figure \ref{fig:fig2}(a), we therefore insert the TOF between two ports of a commercial fused fiber coupler with a 50:50 coupling ratio (Thorlabs FC780-50B-FC) to form a NFS ring resonator that is reasonably close to being critically coupled, albeit with very low finesse. The nanofiber segment is enclosed in a vacuum system filled with warm Rb vapor, with the 50:50 coupler and the majority of the fiber ring outside of the vacuum system.

We utilize the $5^{2}S_{1/2} (F = 2) \rightarrow 5^{2}P_{3/2} (F' = 1-3)$ $^{85}$Rb transition shown in Figure \ref{fig:fig2}(b) and record transmission spectra as the probe frequency $\omega_{p}$ is scanned through the atomic resonance $\omega_{a}$. For the transmission spectrum shown in Figure \ref{fig:fig2}(c), the TOF is temporarily disconnected from the ring cavity allowing a single-pass measurement through the nanofiber segment. The Doppler and transit-time broadened absorption profile in Figure \ref{fig:fig2}(c) shows the interaction of the nanofiber evanescent mode with the Rb atoms \cite{spillane2008}, and a single-pass optical depth in our system of OD $\sim 0.5$ on Rb resonance ($\Delta_{ap} = 0$). 

For the data shown in Figure \ref{fig:fig2}(d), the TOF is re-connected in the ring cavity geometry and the measurement is repeated. The cavity shows a FSR $(76 \pm 12)$ MHz and a measured finesse of $\mathcal{F} =3.6 \pm 0.4$  away from Rb resonance  (near $\Delta_{ap} \sim -1$ GHz). At this detuning, the cavity is close to critical coupling (note that the transmission does not drop all the way to zero on cavity resonance due to a slight mismatch between the coupling loss and the intrinsic internal loss, as well as small polarization errors due to uncompensated birefringence in the nanofiber segment). The main result is seen as the detuning is scanned through $\Delta_{ap} = 0$, where the increasing absorption by the Rb atoms provides an additional internal cavity loss mechanism that moves the cavity farther from critical coupling into the strongly undercoupled regime \cite{choi2001}.

\begin{figure}[b]
\includegraphics[width=3.25in]{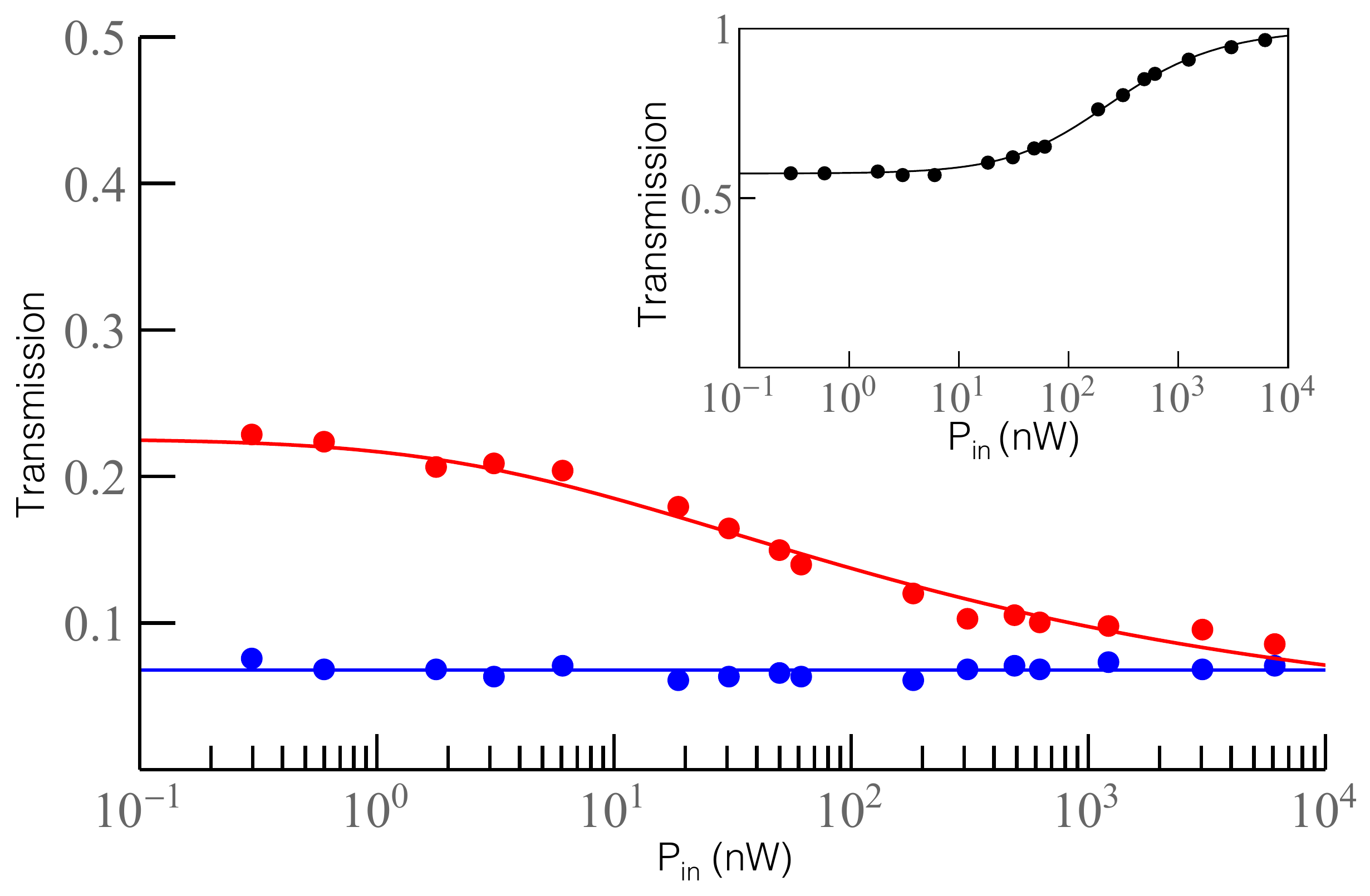}
\vspace*{-.1in}
\caption{(Color online) Nonlinear response of the atom-cavity system.  The main plot shows transmission of two different cavity resonances as a function of input power:  ``with atoms''   (red curve; $\Delta_{ap}  \sim 0$), and ``without atoms''   (blue curve; $\Delta_{ap} \sim -6$ GHz).  ``With atoms'', the cavity moves from being undercoupled (trans. $\sim 23\%$) back to near-critically coupled (trans. $\sim 8\%$) due to saturation of Rb loss at higher input powers. For reference, the boxed inset shows saturation of Rb loss in a single-pass through the nanofiber segment.  All curves in the figures are fits to the data using a simple nonlinear transmission model \protect\cite{jones2014}.}
\label{fig:fig3}
\end{figure}

Figure \ref{fig:fig3} demonstrates the strong nonlinear response of this atom-cavity system. The main plot shows transmission of a cavity resonance ``with atoms'' (red curve; $\Delta_{ap}  \sim 0$)  as a function of input power $P_{in}$. As $P_{in}$ is increased, the Rb excited state population begins to saturate \cite{jones2014}, resulting in a reduction of atomic absorption loss and a return to near critical coupling when the absorption is fully saturated at higher powers. In contrast, the transmission of a cavity resonance ``without atoms'' (blue curve; $\Delta_{ap} \sim -6$ GHz ) is linear with $P_{in}$. 

For reference, the inset to Figure \ref{fig:fig3} shows single-pass transmission through the nanofiber segment (ie. with the TOF temporarily disconnected from the ring cavity) over the same range of input power. The onset of absorption saturation at remarkably low  powers ($P_{in} \sim 10$ nW) is fundamentally due to the small mode area guided by the nanofiber segment \cite{jones2014} and is the origin of the nonlinearity in the main plot.   With a finesse of only  $\mathcal{F} \sim 3.6$, the cavity field buildup factor is on the order of unity, and the rapid change in cavity transmission (red curve) also begins around $P_{in} \sim 10$  nW.

Next we consider the potential of the NFS ring resonator geometry for cavity QED applications using the pioneering work of Kato and Aoki as a benchmark \cite{kato2015}. There, the strong coupling regime was reached using a trapped cesium atom in a nanofiber Fabry-Perot cavity with an overall length of 33 cm, FBG mirror reflectivity of 99.5\%, and a $\mathcal{F} \sim 40$ \cite{kato2015}. The key to the strong coupling was high-quality atom trapping near the nanofiber surface \cite{vetsch2010}, and the remarkably small mode volume of the all-fiber cavity \cite{wuttke2012}. In the NFS ring resonator geometry considered here, comparable mode volumes are easily achievable, and comparable finesse values are within reach given that conventional fiber ring resonators with finesse values of several hundred have been realized \cite{ma2015} and ultrahigh transmission (99.95\%) TOFs  have been achieved \cite{hoffman2014}.  Consequently, the NFS ring resonator may provide an alternative cavity QED platform by adopting atom trapping strategies that are compatible with a ring cavity geometry.

In summary, we have described an atom-cavity system comprised of a nanofiber-segement (NFS) ring resonator and warm Rb vapor.  We observed a strong interaction between the cavity field and the atoms resulting in nonlinear transmission at ultralow (nW) power levels. This  ability to optically control the internal loss of a ring resonator system could form the basis for a number of low-power, but slow (narrowband), switching and modulation technologies \cite{heebner1999,choi2001}.  The all-pass geometry utilized here could be easily changed into a drop-add configuration by adding a second coupling fiber, and the possibility of having multiple nanofiber segments within the ring is also interesting. The NFS ring resonator may also be useful within the context of all-fiber cavity QED \cite{kato2015,wuttke2012}.

This work was supported by the National Science Foundation under grant No. 1402708, and the Office of Naval Research under grant N00014-15-1-2229.


\end{document}